**TITLE**

An unbiased metric of antiproliferative drug effect *in vitro*

**AUTHORS WITH AFFILIATIONS**

Leonard A. Harris[1,2,5], Peter L. Frick[1,2,5], Shawn P. Garbett[1,2], Keisha N. Hardeman[1,2], B. Bishal Paudel[1,2], Carlos F. Lopez[1,3,4], Vito Quaranta[1,2] and Darren R. Tyson[1,2]

[1]Department of Cancer Biology, Vanderbilt University School of Medicine, Nashville, Tennessee, USA

[2]Center for Cancer Systems Biology at Vanderbilt, Vanderbilt University School of Medicine, Nashville, Tennessee, USA

[3]Department of Biomedical Informatics, Vanderbilt University School of Medicine, Nashville, Tennessee, USA

[4]Department of Biomedical Engineering, Vanderbilt University, Nashville, Tennessee, USA

[5]These authors contributed equally to this work.

Correspondence should be addressed to Darren R. Tyson (darren.tyson@vanderbilt.edu).

**ABSTRACT**

*In vitro* cell proliferation assays are widely used in pharmacology, molecular biology, and drug discovery. Using theoretical modeling and experimentation, we show that current antiproliferative drug effect metrics suffer from time-dependent bias, leading to inaccurate assessments of parameters such as drug potency and efficacy. We propose the drug-induced proliferation (DIP) rate, the slope of the line on a plot of cell population doublings versus time, as an alternative, time-independent metric.

**MAIN TEXT**

Evaluating antiproliferative drug activity on cells *in vitro* is a widespread practice in basic biomedical research[1-3] and drug discovery[4-6]. Typically, quantitative assessment relies on constructing dose–response curves[7] (Supplementary Note and Supplementary Fig. 1). Briefly, a drug is added over a range of concentrations and the effect on the cell population is quantified with a metric of choice[8]. The *de facto* standard metric is the number of viable cells 72 h after drug addition[4,6,8,9]. Being a single-time-point measurement, we refer to this as a "static" drug effect metric. The data is then fit to the Hill equation[10], a four-parameter log-logistic function, to produce a sigmoidal dose–response curve that summarizes the relationship between drug effect and concentration. Parameters extracted from these curves include the maximum effect ($E_{max}$), the half-maximal effective concentration ($EC_{50}$), the half-maximal inhibitory concentration ($IC_{50}$), area under the curve (AUC), and activity area (AA)[4,6,8,9] (Supplementary Fig. 1 and Supplementary Table 1). These are useful for quantitatively comparing various aspects of drug activity across drugs and cell lines.

We contend that dose–response curves constructed using current standard metrics of drug effect can result in erroneous and misleading values of drug-activity parameters, skewing data interpretation. This is because they suffer from time-dependent bias, i.e., the metric value varies with the time point chosen for experimental measurement. We identify two specific sources of time-dependent bias: (i) exponential growth, and (ii) delays in drug effect stabilization. The former can lead to erroneous conclusions, e.g., that a drug is increasing in effectiveness over time, while the latter requires shifting the window of evaluation to only include data points after stabilization has been achieved (Supplementary Fig. 2).

To overcome this problem of bias, we propose as an alternative drug effect metric the drug-induced proliferation (DIP) rate[11,12], defined as the steady-state rate of proliferation of a cell population in the presence of a given concentration of drug. Previously, with related approaches, we quantified clonal fitness[12] and heterogeneous single-cell fates[11] within cell populations responding to perturbations. Here, we show that DIP rate is an ideal metric of antiproliferative drug effect because it naturally avoids the bias afflicting traditional metrics, it is easily quantified as the slope of the line on a plot of cell-population doublings versus time (Supplementary Fig. 2), and it is interpretable biologically as the rate of regression or expansion of a cell population.

To theoretically illustrate the consequences of time-dependent bias in standard drug effect metrics, we constructed a simple mathematical model of cell proliferation that exhibits the salient features of cultured cell dynamics in response to drug (Online Methods, Supplementary Note, Supplementary Fig. 3, and Supplementary Table 2). The model assumes that cells experience two fates, division and death, and that the drug modulates the difference between the rates of these two processes, i.e., the net rate of proliferation. Drug action may occur immediately or gradually over time, depending on the chosen parameter values. In all cases, a stable DIP rate is eventually achieved, and when calculated over a range of drug concentrations a sigmoidal dose–response relationship emerges (Supplementary Note and Supplementary Fig. 3).

We model three scenarios: treatment of a fast-proliferating cell line with a fast-acting drug (Fig. 1a), a slow-proliferating cell line with a fast-acting drug (Fig. 1b), and a fast-proliferating cell line with a delayed-action drug (Fig. 1c). In each case, we generate simulated growth curves in the presence of increasing drug concentrations (Fig. 1, columns 1 and 2) and from these produce static dose–response curves by taking cell counts at single time points between 12h and 120h (Fig. 1, column 3). As expected, in each scenario the shape of the dose–response curve varies depending on the time of measurement. Consequently, parameter values extracted from these curves ($EC_{50}$ and AA) also vary (Fig. 1, columns 4 and 5). Similar results are obtained for an alternative drug effect metric proposed by the U.S. National Cancer Institute's Developmental Therapeutics Program[13] (Supplementary Note and Supplementary Fig. 4). In contrast, DIP rate, being the slope of a line, is independent of measurement time. Using it as the drug effect metric gives a single dose–response curve (Fig. 1, columns 3 and 6) and single values of the extracted drug-activity parameters (Fig. 1, columns 4 and 5).

As a first confirmation of our theoretical findings, we subjected triple-negative breast cancer cells (MDA-MB-231) to the metabolic inhibitors rotenone (Fig. 2a) and phenformin (Fig. 2b). Using fluorescence microscopy time-lapse imaging[11,12,14] (Online Methods), we quantified changes in cell number over time for a range of drug concentrations. For both drugs, we observe a rapid stabilization of the drug effect (<24h delay) and stable exponential proliferation thereafter, reminiscent of the growth dynamics of the theoretical cell lines treated with fast-acting drugs (Fig. 1a,b). We generated dose–response curves from these data using the standard static effect metric and DIP rate for various drug exposure times. Consistent with our theoretical results, the shape of the static-based dose–response curve strongly depends on the time point at which cell counts are taken, an illustration of time-dependent bias. The DIP rate, on the other hand, is free of time-dependent bias and produces a single dose–response curve in both cases.

These DIP rate-based dose–response curves produce interesting insights (Fig. 2a,b). For example, they indicate that while rotenone is much more potent than phenformin ($EC_{50} \cong 8.5$ nM versus 25 $\mu$M), phenformin is more effective ($E_{max}/E_0 \cong -0.1$ versus 0.1). The ordering of potencies (rotenone >> phenformin) could have been garnered from the static dose–response curves, but not the ordering of efficacies, i.e., the static drug effect metric obscures the crucial fact that at saturating concentrations phenformin is cytotoxic (causes cell population regression) while rotenone is partially cytostatic (cell populations continue to expand slowly). This information is obviously critical to studies assessing drug mechanism of action. This example illustrates the perils of biased drug effect metrics and the ability of DIP rate to produce reliable dose-response curves from which accurate quantitative and qualitative assessments of antiproliferative drug activity can be made.

To illustrate the confounding effects that a delay in the stabilization of the drug effect can have, we examined single-cell derived clones of the lung cancer cell line PC9, which is known to be hypersensitive to erlotinib[15], an epidermal growth factor receptor (EGFR) kinase inhibitor. Consistent with our previous report[11], three drug-sensitive PC9-derived clones (DS3, DS4, DS5) each respond to 3 $\mu$M erlotinib with nonlinear growth dynamics over the first 48–72h, followed by stable exponential proliferation thereafter (Fig. 2c). These dynamics are reminiscent of those for the theoretical fast-proliferating cell line with a delayed-action drug (Fig. 1c). Due to the delay in drug action, all three clones have nearly identical population sizes 72h after drug addition for all concentrations considered. The static 72h metric thus produces essentially identical dose–response curves for all clones (Supplementary Fig. 5). In contrast, dose–response curves based on DIP rate make a clear distinction between the clones in terms of their long-term response to drug, i.e., erlotinib is cytotoxic (negative DIP rate) for two of the clones but partially cytostatic (positive DIP rate) for the other (Fig. 2c).

We then investigated the effects of erlotinib and lapatinib (a dual EGFR/human EGFR 2 (HER2) kinase inhibitor) on HER2-positive breast cancer cells (HCC1954; delay ~48h; Fig. 2d). In each case, DIP rate-based dose–response curves produce $EC_{50}$ values more than five-fold larger than their static counterparts, i.e., by the static drug effect metric the drugs appear significantly more potent than they actually are. Taken together with the PC9 results (Fig. 2c), these data illustrate the importance of accounting for delays in drug action when assessing antiproliferative drug activity and further emphasize the ability of the DIP rate metric to produce accurate drug-activity parameters and qualitative conclusions about drug-response dynamics.

Within the last several years, a number of studies have been published reporting drug responses for hundreds of cell lines derived from various cancer types[4,6,9,16,17] and organ sites[8,18,19]. Raw data are available for the responses of over 1000 cancer cell lines to a panel of 24 drugs in the Cancer Cell Line Encyclopedia (CCLE)[6] and for over 1200 cell lines treated with 140 drugs in the Genomics of Drug Sensitivity in Cancer (GDSC) project[9]. These data are largely based on static measurements of cell number after 72h of drug exposure, a metric that we have shown here contains time-dependent bias.

To investigate bias in these datasets, we treated four BRAF[V600E or D]-expressing melanoma cell lines with various concentrations of the BRAF-targeted agent PLX4720, an analog of vemurafenib. We produced experimental growth curves (Fig. 3a), static- and DIP rate-based dose–response curves (Fig. 3b), and extracted $IC_{50}$ values for each cell line and compared these to $IC_{50}$ values obtained from the CCLE and GDSC data sets (Fig. 3c). In all cases, our $IC_{50}$ values correspond closely to the value from at least one of the public data sets. While in three cases the static- and DIP rate-based $IC_{50}$ values correspond within an order of magnitude, in one case (A375) they differ by nearly two orders of magnitude. This discrepancy can be traced to a period of complex, non-linear dynamics (brief regression followed by rebound) observed for this cell line between 24h and 72h post-drug addition (Fig. 3a). This result is particularly intriguing because it shows that, based on DIP rate, this cell line is not much different in terms of drug sensitivity than the other three. Using the biased static drug effect metric, however, one would be led to the incorrect conclusion that it is significantly more sensitive. It is very likely that cases like this abound within these and other similar data sets[16,17] and illustrates the critical need for new antiproliferative drug effect metrics.

Current protocols for cell proliferation assays are based on informal 'rules of thumb', for example, counting cells after 72 h of treatment to ameliorate the impact of complex dynamics and delays in drug response. However, these de facto standards have no theoretical basis and, as demonstrated here, they suffer from time-dependent bias that leads to erroneous conclusions. In light of the widespread applications of cell proliferation assays in oncology, pharmacology, and basic biomedical science[20] (for example, to assess activity of cytokines, cell surface receptors, altered signaling pathways, gene overexpression and silencing, or cell metabolic adaptation to varied microenvironmental conditions), it is imperative that the quality of the metric for antiproliferative assays be improved. Toward this end, we have proposed DIP rate as a viable, unbiased alternative. DIP rate overcomes time-dependent bias by log-scaling cell count measurements to account for exponential proliferation and shifting the time-window of evaluation to accommodate lag in the action of a drug, changes that do not substantially alter experimental design (Supplementary Note and Supplementary Figs. 6–9). Moreover, DIP rate is an intuitive, biologically interpretable metric, with a sound basis in theoretical population dynamics, which faithfully captures, within a single value, the long-term effect of a drug on a cell population. Improving the quality of the metric used in antiproliferative assays has the potential to improve success rates in drug discovery and yield more robust gene-drug associations in biomarker discovery studies.

**ACKNOWLEDGMENTS**

We are grateful to R. Feroze, J. Hao, K. Jameson, and C. Peng for support in experimental data acquisition, to J. Guinney, T. de Paulis, and J.R. Faeder for critical reviews of the manuscript, to W. Pao (Vanderbilt University, Nashville, TN) for providing us the PC9 cell line and to Meenhard Herlyn (Wistar Institute, Philadelphia, PA) for providing the WM115 cell line. This work was supported by: Uniting Against Lung Cancer 13020513 (DRT); Vanderbilt Breast Cancer SPORE (Pilot Project, DRT); US National Institutes of Health U54 CA113007 (VQ), CAU01174706 (VQ), and UL1TR000445 (DRT, VQ). Its contents are solely the responsibility of the authors and do not necessarily represent official views of the National Cancer Institute, the National Center for Advancing Translational Sciences or the National Institutes of Health.

**AUTHOR CONTRIBUTIONS**

DRT, LAH, PLF, SPG conceived and designed the study; DRT, LAH, SPG built the mathematical models and performed simulations; BBP, KNH, PLF acquired experimental data; DRT, LAH, PLF, SPG analyzed and interpreted the experimental data; CFL, DRT, LAH, PLF, SPG, VQ wrote, reviewed, and/or revised the manuscript; CFL, DRT, VQ supervised the study.

**REFERENCES FOR MAIN TEXT**

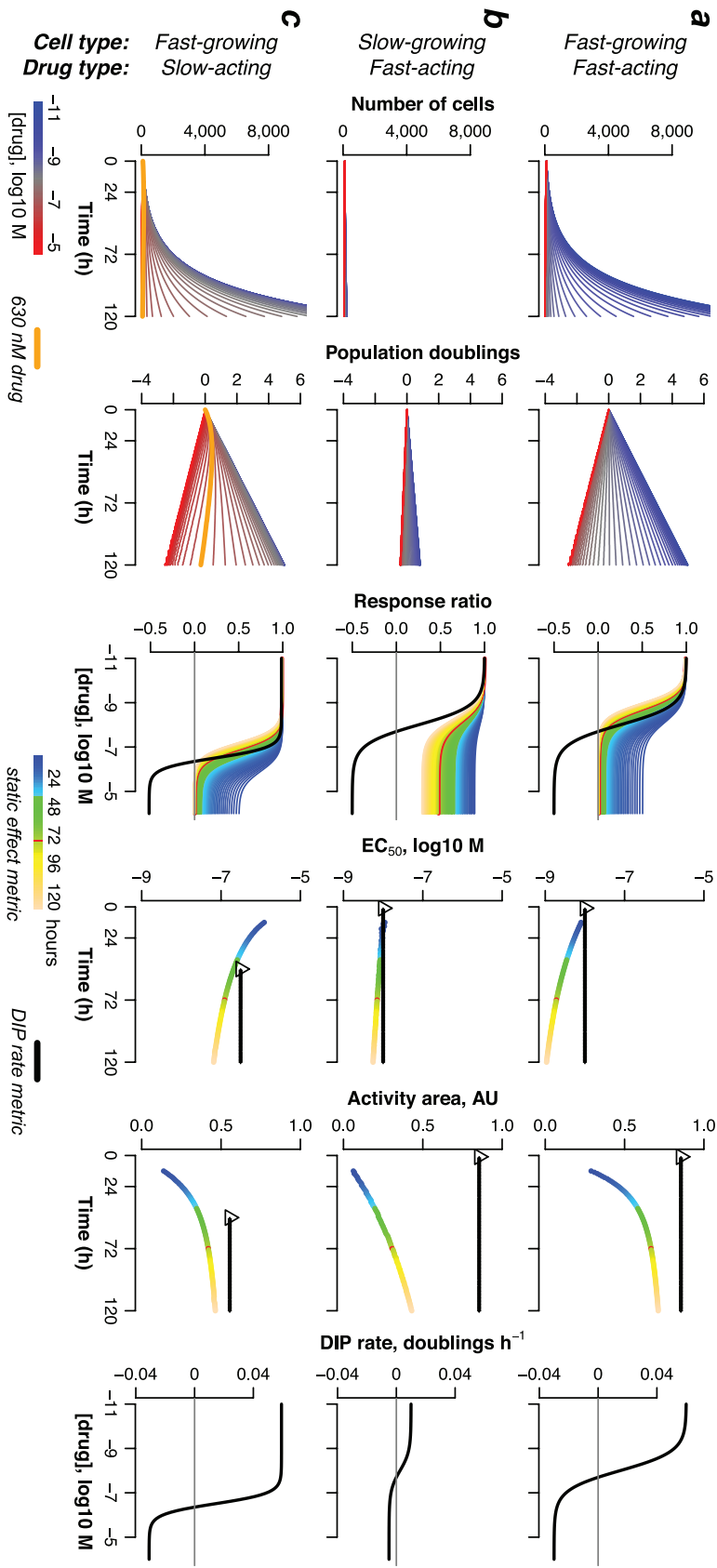

**Figure 1. Theoretical illustration of bias in dose–response curves based on static metrics of drug effect.** Computational simulations of the effects of drugs on: **(a)** a fast-growing cell line treated with a fast-acting drug; **(b)** a slow-growing cell line treated with a fast-acting drug; **(c)** a fast-growing cell line treated with a slow-acting drug. In all cases, *in silico* growth curves, plotted in linear (*column 1*) and $\log_2$ (*column 2*) scale, are used to generate static- (*column 3*) and DIP rate-based (*columns 3 and 6*) dose–response curves, from which values of $EC_{50}$ (*column 4*) and activity area (AA; *column 5*) are extracted. For DIP rate-based values of $EC_{50}$ and AA, the black triangle denotes the first time point used to calculate the DIP rate (i.e., after the drug effect has stabilized; see Online Methods); the black dashed line signifies that the value remains constant for all subsequent time points. Note that the "response ratio" (*column 3*) and "direct effect" (*column 6*) versions of the DIP rate-based dose–response curves (Supplementary Fig. 1) convey complementary information about the activity of a drug on a cell line (see Supplementary Note for further discussion).

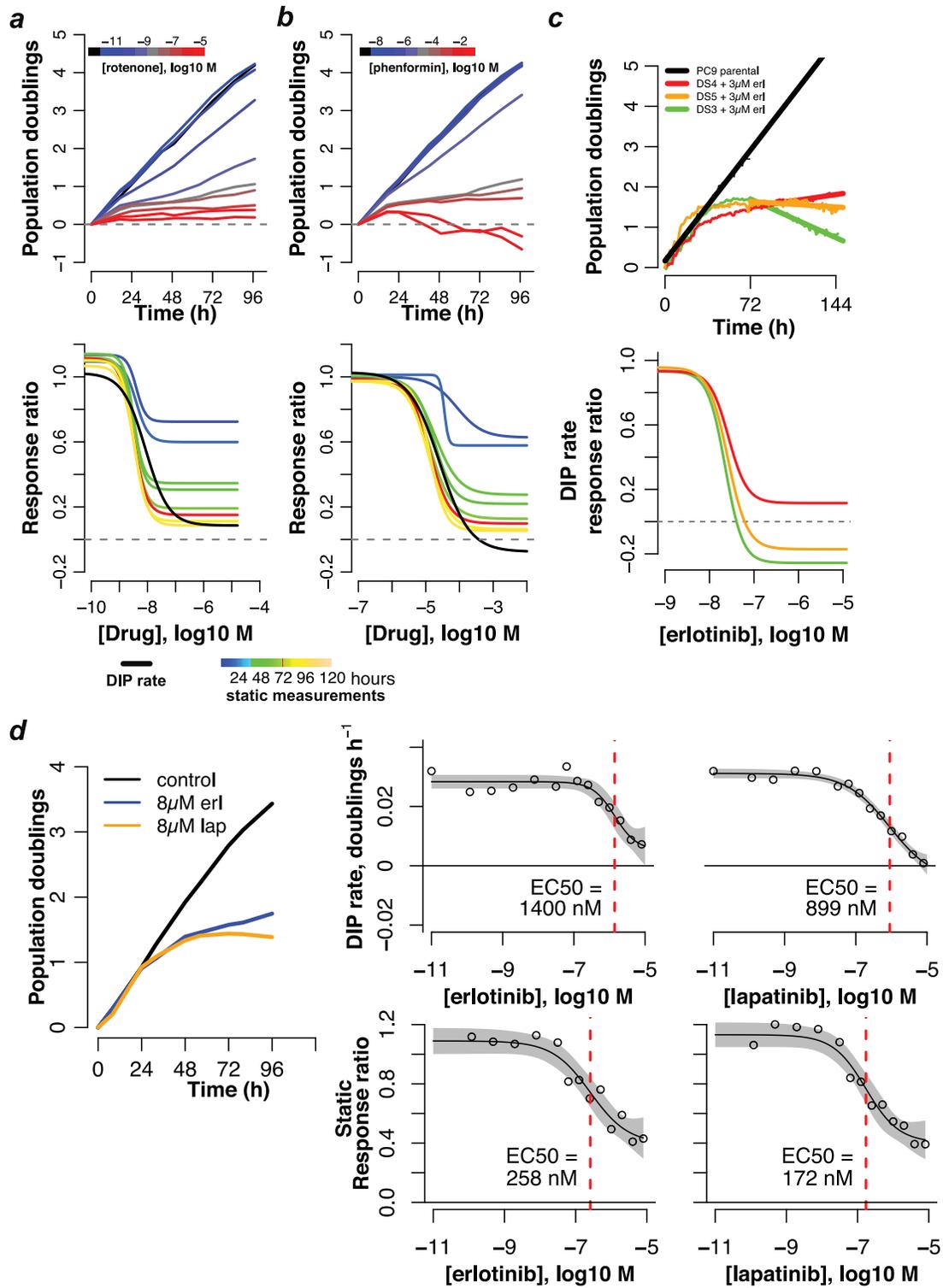

**Figure 2. Experimental illustration of time-dependent bias in dose–response curves for drug-treated cancer cells.** Population growth curves ($\log_2$ scaled) and derived dose–response curves (static- and/or DIP rate-based) for **(a)** MDA-MB-231 triple-negative breast cancer cells treated with rotenone; **(b)** MDA-MB-231 cells treated with phenformin; **(c)** three single-cell-derived drug-sensitive (DS) clones of the EGFR mutant-expressing lung cancer cell line PC9 treated with erlotinib; **(d)** HCC1954 HER2-positive breast cancer cells treated with erlotinib and lapatinib. Data for (*a*) and (*b*) are from single experiments with technical duplicates; data in (*c*) are from individual wells for two experiments containing technical duplicates (growth curves) and from a single experiment with technical duplicates (dose–response curves); data in (*d*) are sums of technical duplicates from a single experiment (growth curves) and mean values (circles) with 95% confidence intervals (gray shading) on the log-logistic model fit (dose–response curves; n=4, 6 for erlotinib and lapatinib, respectively).

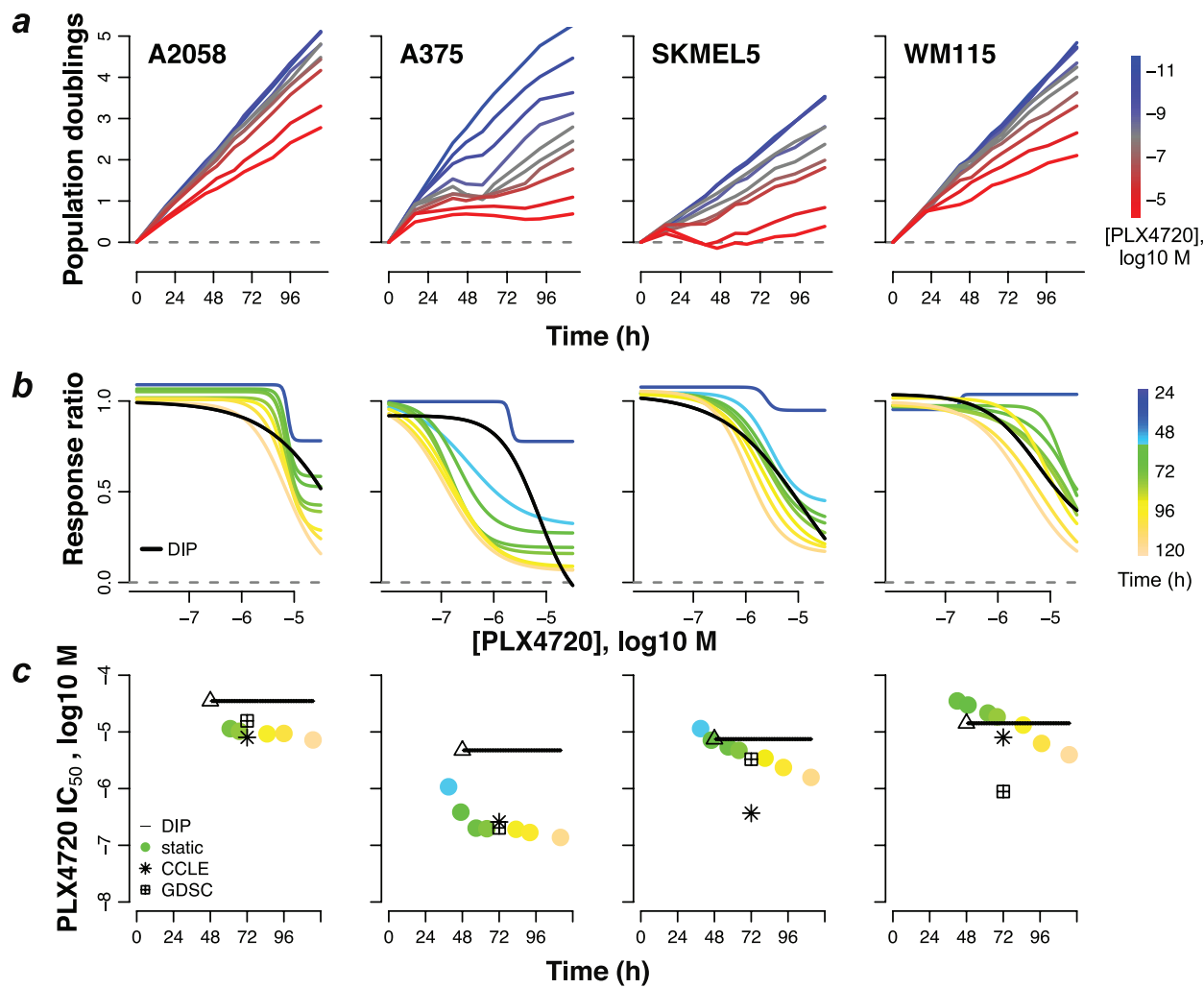

**Figure 3. Bias in potency metrics from publicly available data sets. (a)** Population growth curves (log$_2$ scaled) for four select BRAF-mutant melanoma cell lines treated with various concentrations of the BRAF inhibitor PLX4720; **(b)** dose–response curves based on the static effect metric (colored lines) and DIP rate (black line); (**c**) static- (circles) and DIP rate-based (triangle+line) estimates of $IC_{50}$ for each measurement time point. $IC_{50}$ values obtained from public data sets (CCLE: Cancer Cell Line Encyclopedia; GDSC: Genomics of Drug Sensitivity in Cancer), based on the static 72h drug effect metric, are included for comparison. The triangle denotes the first time point used in calculating the DIP rate and the black line signifies that the value remains constant for all subsequent time points. Data shown are from a single experiment with technical duplicates. Experiment has been repeated at least twice with similar results.

**ONLINE METHODS**

**Dose–response curve fitting**

All drug-response data (theoretical and experimental) were fit with a four-parameter log-logistic function (Supplementary Note) using nonlinear least-squares regression[21] within the R statistical programming environment (http://R-project.org). Fitting was performed using the `drm` function of the `drc` R library[22]. Ninety-five percent confidence intervals for each parameter were obtained using the delta method assuming asymptotic variance[21], as implemented within the `confint` function of the `stats` R library. $EC_{50}$ is a fit parameter of the model. $IC_{50}$ is the concentration at which $E_{drug} = E_0 / 2$ (Supplementary Fig. 1 and Supplementary Table 1), independent of the value of $E_{max}$, and is obtained using the `ED` function of the `drc` library. Activity area (AA; Supplementary Fig. 1) is calculated as

$$\mathrm{AA} = -\sum_{i=1}^{N}\big(E_{drug,i}/E_0 - 1\big)/N \tag{1}$$

where $E_{drug,i}$ is the value of the effect metric at the $i$-th drug concentration and $N$ is the total number of concentrations considered.

**A simple two-state model of drug action on an exponentially proliferating cell population**

We assume that cells can exist in two states, a "no-drug" and a "drug-saturated" state, and that cells in each state can experience two fates, division and death, with kinetic rate constants that are characteristic of the state, i.e., reflecting the effect of the drug (visual representation of the model is provided in Supplementary Fig. 3a). In the presence of drug, cells can transition from the no-drug to the drug-saturated state at a rate proportional to the concentration of drug. Reverse transitions occur at a rate independent of drug. If *Cell* is the number of cells in the no-drug state and *Cell** is the number of cells in the drug-saturated state, then the temporal dynamics of the drug-treated cell population is described by the following pair of coupled ordinary differential equations,

$$\frac{dCell}{dt} = (k_{div} - k_{death} - k_{on} Drug) \cdot Cell + k_{off} \cdot Cell^* \qquad (2)$$

$$\frac{dCell^*}{dt} = (k_{div^*} - k_{death^*} - k_{off}) \cdot Cell^* + k_{on} \cdot Drug \cdot Cell \qquad (3)$$

where $k_{div}$ ($k_{div^*}$) and $k_{death}$ ($k_{death^*}$) are the rate constants for cellular division and death, respectively, in the no-drug (drug-saturated) state, *Drug* is the drug concentration, $k_{on}$ is the rate constant for the transition from the no-drug to the drug-saturated state, and $k_{off}$ is the rate constant for the reverse transition.

At a given drug concentration (assumed to be constant, i.e., drug is neither consumed, removed, nor degraded), a population of cells will eventually reach a dynamic equilibrium in terms of the number of cells in each state. The effective DIP rate of a cell population is then the weighted average of the net proliferation rates (i.e., the difference between the division and death rate constants) of the two individual states (Supplementary Fig. 3b). With increasing drug concentration, the equilibrium shifts increasingly towards the drug-saturated state, asymptotically approaching 100% occupancy. The result is a sigmoidal dose–response relationship between DIP rate and drug concentration (Supplementary Fig. 3c,d). If the values of the rate constants governing the interconversion between the no-drug and drug-saturated state ($k_{on}$ and $k_{off}$) are "large" (effectively infinite), then the dynamic equilibrium between states is achieved immediately upon drug addition. This is known as the partial equilibrium assumption (PEA)[23,24]. Mathematically, the PEA asserts that

$$k_{on} \cdot Drug \cdot Cell = k_{off} \cdot Cell^* \qquad (4)$$

Under this assumption, an analytical solution for the total number of cells, $Cell_T = Cell + Cell*$, can be obtained as a function of time,

$$\ln \frac{Cell_T(t)}{Cell_T(0)} = \frac{k_{off}/k_{on}(k_{div}-k_{death})+Drug(k_{div}^*-k_{death}^*)}{k_{off}/k_{on}+Drug} \cdot t \qquad (5)$$

where $Cell_T(0)$ is the initial number of cells. All theoretical results shown in Figure 1a,b were obtained using equation (5). For the results in Figure 1c and Supplementary Figure 4, numerical integration of equations (2) and (3) was necessary since the values of $k_{on}$ and $k_{off}$ were set such that the PEA does not hold (Supplementary Table 2), i.e., there is a delay in the stabilization of the drug effect. Numerical integration was performed in R using the `deSolve` package[25]. For further details of the model, see Supplementary Note; for all parameter values used in this work, see Supplementary Table 2.

### Cell lines

The PC9 cell line was originally obtained from William Pao (Vanderbilt University). WM115 cells were from Meenhard Herlyn (Wistar Institute). All other cell lines were obtained from the American Type Culture Collection (www.atcc.org). All cell lines are regularly tested for mycoplasma using a PCR-based method (MycoAlert, Lonza, Allendale, NJ) and any positive cultures are immediately discarded. Cell line authentication is provided by ATCC. Authenticity of PC9 and WM115 have not been verified.

### Time-lapse fluorescence microscopic imaging

Time-lapse fluorescence microscopy of cells expressing histone H2B conjugated to monomeric red fluorescent protein (H2BmRFP) to facilitate automated image analysis for identifying and quantifying individual nuclei was performed as previously described[11,12,14].

Briefly, cells are engineered to express H2BmRFP using recombinant, replication-incompetent lentiviral particles and flow sorted for the highest 20% intensity. Cells are seeded at ~2,500 cells per well in 96-well imaging microtiter plates (BD Biosciences) and fluorescent nuclei are imaged using a BD Pathway 855 with a 20× objective in 3×3 montaged images per well at ~15 min intervals for 5 days. Alternatively, fluorescent cell nuclei are imaged twice daily using a Synentec Cellavista High End with a 20× objective and tiling of nine images. DIP rate-based dose–response curves shown in Figure 2c were generated from a single experiment performed at the Vanderbilt High-Throughput Screening Core on a Molecular Devices ImageXpress using similar imaging parameters. The experiment had two technical replicates per condition and images were obtained at 0, 24, 48, 52, 56, 60, 64, 68, 72, 76, 80, 84, 88, 92, 96,100,104,108, and 112 hours after addition of erlotinib at each of eight different concentrations or dimethyl sulfoxide (DMSO) control.

**Other statistical considerations and code availability**

Estimates of DIP rate are determined within an experiment using the sum of cells across all technical replicates at a given time point and obtaining the slope of a linear model of `log2(cell number) ~ time` for time points greater than the observed delay. Minimum delay time is estimated by visual inspection of log-growth curves for the time at which they become approximately linear (for an automated method of estimating the stabilization time point, see Supplementary Note and Supplementary Figs. 6 and 7). All data analysis was performed in R (version 3.2.1) and all raw data and R analysis code is freely available at github.com/QuLab-VU/DIP_rate_NatMeth2016.

**Publicly available data sets**

Drug-response data were obtained from the Genomics of Drug Sensitivity in Cancer (GDSC) project[4,9] website at ftp://ftp.sanger.ac.uk/pub/project/cancerrxgene/releases/release-

[5.0/gdsc_drug_sensitivity_raw_data_w5.zip](5.0/gdsc_drug_sensitivity_raw_data_w5.zip) and from the Cancer Cell Line Encyclopedia

(CCLE)[6] website at [http://www.broadinstitute.org/ccle/](http://www.broadinstitute.org/ccle/) in the data file

`CCLE_NP24.2009_Drug_data_2015.02.24.csv` (user login required).

**Methods-only References**

**SUPPLEMENTARY INFORMATION:** "An unbiased metric of antiproliferative drug effect *in vitro*"

**TABLE OF CONTENTS**

**SUPPLEMENTARY NOTE**

**Dose–response curves and the Hill equation**

In its most general form, the Hill equation can be written as

$$y = \frac{C^h}{C^h + x^h},$$ (S1)

where $x$ is the independent variable, $y$ is the dependent variable, $C$ is a constant that when equal to $x$ results in a value of $y = 1/2$, and $h$ is the Hill coefficient (note that $C$ and $h$ can both be positively or negatively valued). For cell proliferation assays, where $E_{drug}$ is the effect induced by drug at concentration $drug$, $E_{max}$ is the maximum achievable drug effect, and $E_0$ is the effect in the absence of drug, dose-response curves can be constructed using equation (S1) with $x$ as the drug concentration, $y$ as the ratio $(E_{drug} - E_{max}) / (E_0 - E_{max})$, the constant $C > 0$ as the "half-maximal effective concentration," denoted as $EC_{50}$, and $h > 0$ (assuming drug inhibits cell population growth), i.e.,

$$\frac{E_{drug} - E_{max}}{E_0 - E_{max}} = \frac{EC_{50}^{\,h}}{EC_{50}^{\,h} + drug^h}.$$ (S2)

Equation (S2) is known as a four-parameter log-logistic function, the four parameters being $E_{max}$, $E_0$, $EC_{50}$, and $h$. Besides these four, additional drug-activity parameters that can be extracted from these curves include $IC_{50}$ (the half-maximal inhibitory concentration), area under the curve (AUC), and activity area (AA; the inverse of AUC)[1-6] (see Supplementary Table 1 for definitions of these and other relevant terms). These parameters can be used to compare various aspects of drug activity quantitatively across drugs and cell lines. We refer to equation (S2) (reproduced in Supplementary Fig. 1a) as the "scaled" form of the dose-response curve because the $y$-axis values are scaled between 0 and 1.

Dose–response curves are generally plotted with drug concentrations along the $x$-axis in $\log_{10}$ scale in order to easily visualize a broad range of concentrations. In this view, we can show that the Hill coefficient $h$ is inversely proportional to the slope of the curve at the $EC_{50}$ by noting that $\log_{10}(x) = \ln(x) / \ln(10)$ and calculating the derivative of equation (S2) with respect to $\log_{10}(x)$,

$$\begin{aligned} \frac{dy}{d\log_{10} x} &= \frac{dy}{dx} \cdot \frac{dx}{d\log_{10} x} \\ &= \frac{-\ln(10) h x^h C^h}{\left(C^h + x^h\right)^2}. \end{aligned}$$ (S3)

Evaluating this at $x = C$ gives

$$\frac{dy}{d\log_{10} x}\Big|_{x=C} = \frac{-\ln(10)}{4} h.$$ (S4)

Hence, the larger the value of $h$ the steeper the dose–response curve at the $EC_{50}$, and vice versa[1].

Furthermore, we define the $IC_{50}$ (half-maximal inhibitory concentration) as the concentration of drug at which $E_{drug} = E_0 / 2$ (Supplementary Table 1). Substituting this into equation (1) of the main text, setting $drug = IC_{50}$, and rearranging gives

$$IC_{50}^h = \left(\frac{0.5 \cdot E_0}{0.5 \cdot E_0 - E_{max}}\right) \cdot EC_{50}^h. \tag{S5}$$

We see, therefore, that if $E_{max} < 0$ then $IC_{50} < EC_{50}$ (Supplementary Fig. 1), and vice versa. Importantly, we also see that $IC_{50}$ is not defined if $E_{max} \geq E_0/2$, which makes intuitive sense. Note that when cell number after a specified period of drug exposure is used as the drug effect metric (the standard "static" metric), $E_{max}$ is always $\geq 0$. However, for dynamic metrics such as DIP rate, $E_{max}$ can be positive or negative.

In practice, $E_0$ and $E_{max}$, along with $EC_{50}$ and $h$, are treated as adjustable parameters that can be estimated based on a numerical fit to experimental drug-response data using nonlinear least-squares regression. To accomplish this, equation (S2) is rearranged to solve for $E_{drug}$, which we refer to as the "direct effect" form of the dose–response curve (Supplementary Fig. 1b). In the main text, we argue that a valuable characteristic of the DIP rate is that it is directly interpretable from a biological perspective. We support this claim by showing direct-effect DIP rate-based dose–response curves in column 6 of Figure 1. These can be used to directly infer the characteristics of the theoretical growth curves (e.g., proliferation rates in the absence of drug, DIP rates at saturating drug concentrations) shown in columns 1 and 2 of Figure 1. In general, we suggest that all DIP-rate based drug-response data be reported and stored (e.g., in public databases) in this form.

In contrast to the DIP rate, static drug effect metrics (e.g., cell number 72h after drug exposure) do not have any biological meaning except with respect to a reference value, such as untreated control. As such, static dose–response curves are usually displayed in terms of the "response ratio" $E_{drug}/E_0$. The equation for this form of the dose–response curve is obtained by simply dividing the direct-effect form by $E_0$ (Supplementary Fig. 1c). In Figures 1–3 of the main text and Supplementary Figures 4 and 5, we show numerous examples of dose–response curves plotted in terms of response ratios, both for traditional drug effect metrics and DIP rate. This allows us to directly compare traditional and DIP rate-based dose–response curves on the same plot. Furthermore, in the case of DIP rate, direct-effect and response-ratio dose–response curves convey complementary, but distinct, information, which may be important for interpreting drug effects within different contexts. For example, in Figure 1, the fast- and slow-proliferating cell lines treated with fast-acting drugs exhibit identical response-ratio dose–response curves (by construction; see Supplementary Table 2). This indicates the same dose-dependent drug activity in these cell lines despite the substantially different basal rates of proliferation and rates of cell loss at high drug concentrations, evident in their direct-effect dose–response curves. When assessing drug activity across cell lines, in particular, both sets of information are important. We refrain, therefore, from advocating for one form of the dose–response curve over the other. Note, however, that response ratios are easily calculated from the information contained within direct-effect dose–response curves, which is why we advocate storing drug-response data in this form.

**Simple two-state model of drug concentration-dependent fractional proliferation**

Substantial evidence exists that anticancer drugs affect cultures of solid tumor-derived cancer cells by both inducing cell death and elongating cell cycle times[7,8]. We refer to the concept that the dynamics of drug-treated cell populations are a combined effect of multiple different cell fates (e.g., cell division, death, and survival) as "fractional proliferation" (see Supplementary Table 1). We previously described a Quiescence-Growth (QG) mathematical model that assumes that drugs can affect the rates of three different cell fates: division, death, and entry into quiescence[8]. We used the model to consolidate drug-induced single cell fate decisions with cell population dynamics for several cell lines treated with a variety of drugs[8]. However, the model does not describe the relationship between the rates of cell fates and the concentration of drug. We therefore modified the QG model, as described in detail below, to provide a mechanistic basis for the observed log-logistic relationship between drug concentration and steady-state rate of cell proliferation (i.e., DIP rate).

As described in Online Methods, we assume that cells can exist in two states: a "no-drug" and a "drug-saturated" state. Cells within each state can divide and die at rates that are characteristic of the state. Furthermore, cells can transition between the two states, with the rate of transition from the no-drug to the drug-saturated state being dependent upon drug concentration (Supplementary Fig. 3a). Since the effect of an anticancer drug is to reduce the net rate of proliferation of a cell population, we impose that the proliferation rate of the no-drug state be larger than that of the drug-saturated state (Supplementary Fig. 3b). Defining *Cell* and *Cell\** to be the populations of cells in the no-drug and drug-saturated states, respectively, the model can be written in kinetic terms as

$$Cell \xrightarrow{k_{div}} Cell + Cell \tag{S6}$$

$$Cell \xrightarrow{k_{death}} \emptyset \tag{S7}$$

$$Cell + Drug \xrightarrow{k_{on}} Cell^* + Drug \tag{S8}$$

$$Cell^* \xrightarrow{k_{off}} Cell \tag{S9}$$

$$Cell^* \xrightarrow{k_{div^*}} Cell^* + Cell^* \tag{S10}$$

$$Cell^* \xrightarrow{k_{death^*}} \emptyset \tag{S11}$$

where $\emptyset$ represents cell death (the null state) and all rate constants are as illustrated in Supplementary Figure 3a.

Assuming continuous and deterministic dynamics, the time course of a drug-treated cell population is described by a coupled set of ordinary differential equations (ODEs), which can be derived directly from reaction set (S6)–(S11),

$$\frac{dCell}{dt} = (k_{div} - k_{death} - k_{on} \, Drug) \cdot Cell + k_{off} \cdot Cell^*, \tag{S12}$$

$$\frac{dCell^*}{dt} = \left(k_{div^*} - k_{death^*} - k_{off}\right) \cdot Cell^* + k_{on} \cdot Drug \cdot Cell. \tag{S13}$$

Equations (S12) and (S13) are presented as equations (2) and (3) in the main text, respectively. In general, these equations must be solved numerically. However, if the rate constants $k_{on}$ and $k_{off}$ that govern the transitions between the no-drug and drug-saturated states are "large" (effectively infinite), then a solution can be obtained analytically under the partial equilibrium assumption[9] (PEA). The PEA amounts to setting the rates of reactions (S8) and (S9) equal to each other, i.e.,

$$k_{on} \cdot Drug \cdot Cell = k_{off} \cdot Cell^*. \tag{S14}$$

Equation (S14) is presented as equation (4) in the main text. If we define the total cell population as

$$Cell_T \equiv Cell + Cell^*, \tag{S15}$$

then we can obtain expressions for the no-drug and drug-saturated cell populations as a function of the total cell population by substituting equation (S15) into equation (S14) and rearranging,

$$Cell = \frac{\frac{k_{off}}{k_{on}}}{\frac{k_{off}}{k_{on}} + Drug} \cdot Cell_T, \tag{S16}$$

$$Cell^* = \frac{Drug}{\frac{k_{off}}{k_{on}} + Drug} \cdot Cell_T. \tag{S17}$$

Summing equations (S12) and (S13) gives a single ODE describing the temporal dynamics of the total cell population,

$$\frac{dCell_T}{dt} = \frac{dCell}{dt} + \frac{dCell^*}{dt} = (k_{div} - k_{death}) \cdot Cell + (k_{div^*} - k_{death^*}) \cdot Cell^*. \tag{S18}$$

Substituting equations (S16) and (S17) into equation (S18) and rearranging gives

$$\frac{dCell_T}{dt} = \frac{\frac{k_{off}}{k_{on}}(k_{div} - k_{death}) + Drug(k_{div^*} - k_{death^*})}{\frac{k_{off}}{k_{on}} + Drug} \cdot Cell_T, \tag{S19}$$

which can be solved analytically by separation of variables,

$$\ln \frac{Cell_T}{Cell_T(0)} = \frac{\frac{k_{off}}{k_{on}}(k_{div} - k_{death}) + Drug(k_{div^*} - k_{death^*})}{\frac{k_{off}}{k_{on}} + Drug} \cdot t. \tag{S20}$$

Equation (S20) is presented as equation (5) in the main text.

The DIP rate is defined as the slope of the line on a semi-log$_2$ plot of cell number vs. time. Therefore, under the PEA, the DIP rate for our model is

$$DIP = \frac{1}{\ln 2} \cdot \frac{\frac{k_{off}}{k_{on}}(k_{div} - k_{death}) + Drug(k_{div^*} - k_{death^*})}{\frac{k_{off}}{k_{on}} + Drug}. \tag{S21}$$

At zero drug concentration, equation (S21) reduces to

$$DIP_0 = \frac{1}{\ln 2} \cdot (k_{div} - k_{death}). \tag{S22}$$

At maximum (infinite) drug concentration, equation (S21) reduces to

$$DIP_{max} = \frac{1}{\ln 2} \cdot (k_{div^*} - k_{death^*}). \tag{S23}$$

With $DIP$, $DIP_0$, and $DIP_{max}$ as $E_{drug}$, $E_0$, and $E_{max}$, respectively, equation (S2) describing the dose-response curve can be rewritten as

$$\frac{DIP - DIP_{max}}{DIP_0 - DIP_{max}} = \frac{EC_{50}^h}{EC_{50}^h + Drug^h}. \tag{S24}$$

Substituting equations (S21)–(S23) into equation (S24) and rearranging gives

$$\frac{\frac{k_{off}}{k_{on}}}{\frac{k_{off}}{k_{on}}+Drug} = \frac{EC_{50}^h}{EC_{50}^h+Drug^h}. \tag{S25}$$

Thus, we see that under the PEA our model predicts a sigmoidal DIP rate-based dose-response curve with

$$EC_{50} = \frac{k_{off}}{k_{on}} \tag{S26}$$

and $h$ = 1 (Supplementary Fig. 3c). We can now obtain an expression for the DIP rate in terms of $DIP_0$, $DIP_{max}$, and $EC_{50}$ by substituting equations (S22), (S23), and (S26) into equation (S21) and rearranging,

$$DIP = \frac{1}{\ln 2} \cdot \frac{EC_{50} \cdot DIP_0 + Drug \cdot DIP_{max}}{Drug + EC_{50}}. \tag{S27}$$

Even in cases where the PEA does not hold, our two-state model (S6)–(S11) predicts a sigmoidal relationship between DIP rate and drug concentration, although not exactly of the Hill form (Supplementary Fig. 3d). The curve must be obtained in this case through numerical integration of equations (S12) and (S13), as an analytical solution is not possible.

**Cell proliferation assay developed by the National Cancer Institute Developmental Therapeutics Program**

An alternative effect metric that is sometimes used in cellular proliferation assays is the difference between cell number after a specified period of drug exposure and cell number at the time of drug addition[1,10-13]. A protocol for generating dose–response curves based on this metric has been developed by the U.S. National Cancer Institute's Developmental Therapeutics Program[10,11] (NCI DTP). The NCI DTP has been performing *in vitro* analyses of therapeutic compounds on a panel of 60 cancer cell lines (the NCI60) for several decades[10]. This program has screened thousands of compounds for their anticancer properties, of which the U.S. Federal Drug Administration has licensed dozens as clinical anticancer agents. The approach for characterizing anticancer drug response uses an indirect assay of cell counts based on spectrophotometric absorbance readings (described in detail at https://dtp.cancer.gov/discovery_development/nci-60/methodology.htm). If the mean absorbance of the treated sample after a specified period of drug exposure, *Ti*, is greater than or equal to the mean absorbance on day zero, *Tz*, then the response ratio is calculated as (*Ti* − *Tz*) / (*C* − *Tz*), where *C* is the absorbance of the control-treated cells at the same time point as *Ti*. However, if *Ti* < *Tz*, then the response ratio is calculated as (*Ti* − *Tz*) / *Tz*. In the terminology used in this paper, *Ti* − *Tz* is the drug effect metric and it is "dynamic" because it is based on measurements at more than one time point (Supplementary Table 1).

The NCI DTP promotes its ability to characterize compounds for their differential or selective patterns of drug sensitivity, generally assessed by metrics of potency, across the panel of cancer cell lines[14]. In early work describing the development of the screening assay methodology, it was noted that dose–response data was strongly time dependent, such that "the magnitude of measured drug sensitivity in a given cell line is primarily dependent upon culture [and drug exposure] duration."[14] However, because the NCI DTP metric is based on direct cell counts (not log scaled) and does not account for delays in the action of the drug, it is subject to the same sources of time-dependent bias that afflict the standard static effect metric (Figs. 1-3 of the main text). To demonstrate this, in Supplementary Figure 4a we provide a visual illustration of the NCI DTP dynamic metric applied to the population growth curve at 630 nM drug concentration from Figure 1c of the main text. We see immediately from this plot the confounding effects that bias can have on this metric: depending on when the cell count measurement is taken, the drug could be interpreted as being partially cytostatic (<96h), fully cytostatic (96h), or cytotoxic (>96h). We further quantify the effects of bias in Supplementary Figure 4b, where we generate dose–response curves over a range of measurement time points. As with the static effect metric (Fig. 1), the shape of the NCI DTP-based curve strongly depends on the chosen measurement time point. In light of this fact, we caution against using the NCI DTP approach to infer specific biological activities of drugs, contrary to recent reports in the literature[1,12,15,16].

**Practical considerations for using the DIP rate metric in high-throughput screening assays**

The following issues, related to both experimental design and data analysis, should be considered when adopting DIP rate as a standard drug effect metric in high-throughput screening assays.

*Duration of the experiment*
The recommended total assay duration is seven days since, in most cases, we observe stabilization of the DIP rate within this time frame. With modern environmentally controlled microscopes, these assays are easily set up and automated with robotics. It is also possible to perform longer-term assays if desired, e.g., to confirm DIP rate stability over longer time scales. Care should be taken, however, that cell density does not exceed 70% well surface area, as this can impact population size-independent drug action.

*Drug additions and media changes*
The procedure in use in our laboratory involves seeding cells into multiwell plates, allowing them to adhere and/or acclimate for 16–24h, and then adding drug-containing fresh medium (defined as time zero). After 72h, medium is replaced with fresh drug-containing medium and the assay is continued for another 72h.

*Range of drug concentrations*
Published $EC_{50}$ or $IC_{50}$ values, if available, are a good starting point for estimating the range of concentrations to be used (despite the potential bias in their values, as illustrated in this paper). A wide range around that value (e.g., four-fold dilutions for a total of eight drug concentrations) should span the relevant range in many cases. It is also possible to use 10-fold dilutions to cover a broader range, e.g., if published $EC_{50}$ or $IC_{50}$ values are not available. To minimize off-target effects, we typically use concentrations below the maximum drug solubility in aqueous buffers.

*Cell counting methodology*
A variety of cell counting methods can be employed. In our experience, direct cell counting from fluorescence microscopy images of cells with labeled nuclei is ideal. Individual nuclei are automatically counted by automated image segmentation from digital time series images stored by a computerized microscope. Other means of labeling cells, or even unlabeled cells, can be used, although this may place more demands on automated cell counting. Indirect cell counting via fluorescence intensity measurements, such as CellTiterGlo, is also possible by implementing a scheme of replicate plates treated in parallel and harvested at different time points.

*Frequency of sampling*

In part, the frequency of cell counting is dictated by throughput (i.e., large numbers of plates may require accommodation of longer handling times). To test the robustness of our automated DIP rate estimation algorithm (see below) to variations in sampling frequency, we calculated stabilization times and DIP rates from the complete DS3 data set (n=113) and from successively subsampled versions of the data set (n=57, 29, 15 and 8; Supplementary Fig. 7). The stabilization times obtained from all of these data sets fell within a 10h window (68−77h). DIP rates for all data sets with ≥15 data points varied by less than 1%; the DIP rate value from the smallest data set (n=8) was still within 15% of the value obtained from the complete data set. These results indicate that obtaining 2−3 images per day (8−12h intervals) is likely sufficient in many cases to obtain reasonable estimates of drug effect stabilization times and DIP rates.

*Aggregation of data from technical and biological replicates*

Technical replicates (multiple wells in a plate) are unnecessary if sufficient numbers of cells can be quantified. However, if technical replicates are generated, the cell counts at each time point should be summed rather than averaged to reduce bias introduced by replicates with fewer cells. For biological replicates (separate repeat experiments), we calculate the mean DIP rate over all replicates.

*Collected data structure*

Cell count data is structured as a seven-column matrix, where the columns are:  (1) time of measurement, (2) cell count, (3) cell line ID, (4) drug name, (5) drug concentration, (6) well number, and (7) date. This structured data can be sent directly to the automatic DIP rate estimation algorithm (next subsection) to generate DIP rate-based dose–response curves.

*Automated estimation and statistical confidence of DIP rate*

As discussed in the main text, DIP rate is defined as the rate of growth of a cell population at steady state, i.e., after the effect of a drug (or any perturbagen) has stabilized. On a plot of cell population doublings vs. time, a stabilized drug effect corresponds to a sustained linearity of the growth curve. In high-throughput screening assays, computational algorithms that can detect drug effect stabilization and calculate DIP rates automatically from cell count data are necessary. To identify the optimal range of data over which to calculate DIP rate, we present here an algorithm using two commonly applied metrics of linear model fitting: adjusted $R^2$ and root-mean-squared error (RMSE). Adjusted $R^2$ quantifies how much the change in cell number can be explained by changes in time, whereas RMSE is a metric of how close the measured cell numbers are to values predicted by the linear model fit[17]. Our approach is to first fit a linear model to all data points from a time course and calculate adjusted $R^2$ and RMSE values. We then progressively remove data points from the beginning of the time course, obtaining new model fits and recalculating the adjusted $R^2$ and RMSE values until five data points remain (for a five-day experiment where cell counts are acquired every 12h, this amounts to about half the data points). DIP rate stabilization can then be defined either as (1) the time point at which the adjusted $R^2$ value is at its maximum, i.e., when time and change in $\log_2$ cell number are most highly correlated, or (2) the time point at which the derivative of the RMSE curve (fit with a polynomial) first reaches zero (within a defined tolerance), i.e., the earliest time point for which RMSE is not significantly improved by exclusion of this point. To demonstrate the effectiveness of this approach, in Supplementary Figure 6 we apply it to the PC9 subline data shown in Figure 2c of the main text. Both the adjusted $R^2$ and RMSE metrics produce similar values of stabilization times and DIP rates for each subline (DS3, DS4, DS5). Stabilization times are within a 10h window for DS5 and a 5h window for DS3 and DS4; DIP rates are within 0.001 doublings $h^{-1}$ for all sublines. Importantly, all PC9 sublines (including others not shown) demonstrate a stabilized DIP rate by 72h, justifying our use of this time point in the analyses presented in the main text. Source code implementing this approach (`dipDRC.r`) is freely available at [github.com/QuLab-VU/DIP_rate_NatMeth2016](github.com/QuLab-VU/DIP_rate_NatMeth2016).

*Variations around a mean cell seeding density and measurement time*

Variations in the number of cells seeded per well are inevitable in high-throughput assays, even when using automated high-precision robotic platforms[18]. Because cell populations grow exponentially, the effects of these variations amplify over time and can significantly affect the calculated values of drug effect metrics and extracted drug-response parameters, such as potency and efficacy. Variability in the time point at which cell count measurements are acquired can have a similar effect. The consequences of these sources of variability can be quantified by considering the basic exponential growth formula,

$$X(t) = X_0 e^{kt}, \tag{S28}$$

where $X(t)$ is the cell count at time $t$, $X_0$ is the initial cell count, and $k$ is the exponential growth rate, which is proportional to the DIP rate $d=k/\ln(2)$ (i.e., DIP rate is defined on a $\log_2$ basis and has units of [doublings/time]). From equation (S28), we see immediately that $k$, and hence $d$, is completely independent of the initial cell count and the measurement time point. *DIP rate is thus unaffected by any variations in these quantities,* assuming that intrinsic stochastic effects are minimal (see below). This quality of DIP rate makes it particularly well suited as a standard drug effect metric.

The same cannot be said for $X(t)$, which we refer to in the main text as the traditional static drug effect metric. If we assume that the initial cell count, $X_0$, is distributed according to a normal (Gaussian) distribution with mean $\mu$ and variance $\sigma^2$, denoted as

$$X_0 \sim \mathcal{N}(\mu, \sigma^2), \tag{S29}$$

and we assume that the sampling time $t$ is a constant $\bar{t}$, then the final cell count is also normally distributed according to

$$X(\bar{t}) \sim \mathcal{N}(e^{k\bar{t}}\mu, e^{2k\bar{t}}\sigma^2). \tag{S30}$$

We see, therefore, that in the absence of variations in the measurement time, variability in the seeding density broadens the distribution of final cell counts by a constant factor that increases exponentially with the intrinsic growth rate of the cell population and with the measurement time point. In practice, $X(t)$ is almost always considered in terms of a ratio with respect to untreated control, a quantity that we refer to as the "response ratio." The control value is often taken as the mean cell count at time $t$ from $N$ (typically <10) control experiments, i.e.,

$$\langle X_c(t) \rangle = \sum_{i=1}^{N} X_c(t)/N, \tag{S31}$$

where the subscript $c$ denotes "control." Assuming that the mean and variance in the seeding density in the control experiments is equivalent to that in drug, we can use equations (S30) and (S31), with $X(\bar{t}) = X_c(\bar{t})$ and $k=k_c$, to derive the distribution

$$\langle X_c(\bar{t}) \rangle \sim \mathcal{N}(e^{k_c\bar{t}}\mu, e^{2k_c\bar{t}}\sigma^2/N). \tag{S32}$$

The response ratio is thus distributed according to the ratio of the two normal distributions in equations (S30) and (S32), which we write as

$$\frac{X(\bar{t})}{\langle X_c(\bar{t}) \rangle} \sim \frac{\mathcal{N}(e^{k\bar{t}}\mu, e^{2k\bar{t}}\sigma^2)}{\mathcal{N}(e^{k_c\bar{t}}\mu, e^{2k_c\bar{t}}\sigma^2/N)}. \tag{S33}$$

An analytical expression for the distribution of the ratio of two normal random variables has been presented by Hinkley[19].

In the case of variability in the measurement time, if we assume that $t$ is normally distributed,

$$t \sim \mathcal{N}(\mu_t, \sigma_t^2), \tag{S34}$$

then the exponential term in equation (S28) is no longer a constant, but is distributed according to a log-normal distribution,

$$e^{kt} \sim \ln \mathcal{N}(k\mu_t, k^2\sigma_t^2). \tag{S35}$$

From equation (S28), the final cell count $X(t)$ is therefore distributed according to the product of a normally distributed random variable (equation S29) and a log-normally distributed random variable (equation S35). This is known as a normal-log-normal (NLN) mixture distribution[20]. Since an analytical expression does not exist for the NLN distribution, we employ numerical simulation to obtain estimates for the distributions of $X(t)$, $\langle X_c(t) \rangle$, and the response ratio.

In Supplementary Figure 8, we show results of our theoretical analysis of the effects of variability in seeding density and measurement time on the traditional static drug effect metric. We use the same two-state model as in the main text (described above) with the "fast proliferating / fast acting" parameter set (Supplementary Table 2). All simulations were performed deterministically using a standard ODE integrator[21]. In Supplementary Figure 8a, we show 100 simulated time courses each for untreated and drug-treated (3 nM) cell populations. Because the simulations are deterministic (no intrinsic noise effects), we normalize the initial cell counts to the mean, i.e., we sample from the distribution

$$X_0 / \mu \sim \mathcal{N}(1, \sigma^2/\mu^2), \tag{S36}$$

with standard deviation $\sigma/\mu$=0.1 (this quantity is also known as the coefficient of variation). To model variations in the measurement time, we sample simulation run times from a normal distribution (equation S34) with mean 72h and standard deviation 5h. We consider two cases: variations in cell seeding density alone and variations in both seeding density and measurement time. In both cases, we perform $10^5$ control simulations (drug=0), divide the final cell counts into $10^4$ groups of 10, and calculate mean values for each group. We then perform $10^4$ drug-treated simulations and take the ratio of all pairs of drug-treated and control values to give us $10^8$ samples of the static response ratio. Boxplots of these ratios are shown in Supplementary Figure 8b. Full distributions of $\langle X_c(t) \rangle$, $X(t)$, and the response ratio for variations in cell seeding density alone are shown in Supplementary Figure 8c, where analytical distributions (equations S30, S32, S33) are overlaid with simulation-based distributions. Simulation-based distributions in the case of variations in both cell seeding density and measurement time are shown in Supplementary Figure 8d.

Overall, we see that normally distributed variations in cell seeding density result in a significant distribution of response ratios and that variations in the measurement time act to skew this distribution towards larger values. As such, identical experimental setups can result in a wide range of response ratios, which are used to generate dose–response curves from which drug-response parameters are extracted. This variability may therefore explain, in part, reported discrepancies among drug-response parameters in different publicly-available datasets[3,4]. As emphasized above and in the main text, DIP rate does not suffer from such variability and, hence, its adoption as a standard drug effect metric may improve congruence among such datasets in the future.

*Variations in mean cell seeding density*

In addition to random variations in cell number around a mean seeding density (previous subsection), the absolute number of seeded cells can also greatly impact the dynamics of cellular proliferation and, hence, the precision and accuracy of measured values of drug effect metrics. The origin of this effect is the stochastic nature of cell fate decisions[22,23], e.g., randomness in times to division and death. Such "intrinsic" stochastic effects are particularly prevalent at low cell numbers and, as a rule of thumb, scale as $1/\sqrt{N}$, where $N$ is the number of cells[24]. While a full theoretical treatment of the role of intrinsic stochasticity in cellular proliferation and its effect on drug effect metrics is outside the scope of this paper, we present in Supplementary Figure 9 results of an experimental investigation of variability in estimated DIP rates over a range of mean seeding densities. We treated populations of a BRAF-mutant melanoma cell line (SKMEL5) with 8 $\mu$M of the BRAF inhibitor PLX4720 for 150h and calculated DIP rates based on all data points $\gtrsim$72h. Four time courses were obtained at each seeding density ranging from 312 to 10,000 cells/well (Supplementary Fig. 9a,b). As expected, the variance in the calculated DIP rates decreases with increasing seeding density (Supplementary Fig. 9c; Levene's test p=0.0082). However, the mean values are statistically indistinguishable across seeding densities (p=0.47). These results indicate that reliable estimates of DIP rate can be obtained even at low seeding densities.

**Protocol for generating dose–response curves using DIP rate as the drug effect metric**

Needed:
- Cells with genetically encoded nuclear label (e.g., H2BmCherry).
- 96-well imaging plates (e.g., BD cat#353219).
- Automated fluorescence microscope with 96-well plate-compatible stage (e.g., Synentec Cellavista, BD Pathway or Molecular Devices ImageXpress).
- Drugs/compounds of interest.

Method:
1) Seed cells at ~2,500 cells per well in 96-well imaging plate and incubate overnight in environmentally controlled incubator.
2) Prepare, in complete culture medium, eight four-fold dilutions of each drug (maximum = 4 μM, minimum = 0.24 nM). If prior knowledge exists of expected $EC_{50}$ or $IC_{50}$ (e.g. $EC_{50}$ >> 250 nM, $EC_{50}$ << 1 nM), adjust the range of concentrations accordingly.
3) Replace medium in imaging plate with drug dilutions and two control wells receiving complete medium alone. Time of initial drug addition = 0.
4) Obtain fluorescence microscopy images at least every 8–12h (hourly for more precise DIP rate estimate).
5) Count cells for each time point and condition using digital image segmentation (e.g., ImageJ, Matlab, Python).

6) Structure data into a matrix containing seven columns: (1) time of measurement, (2) cell count, (3) cell line ID, (4) drug name, (5) drug concentration, (6) well number, and (7) date.

7) Generate DIP rate-based dose–response curve by passing the data structure to `dipDRC.r`, an R function available at [github.com/QuLab-VU/DIP_rate_NatMeth2016](github.com/QuLab-VU/DIP_rate_NatMeth2016) (see `makeDRCexample.r` for an example of usage).

**SUPPLEMENTARY RESULTS**

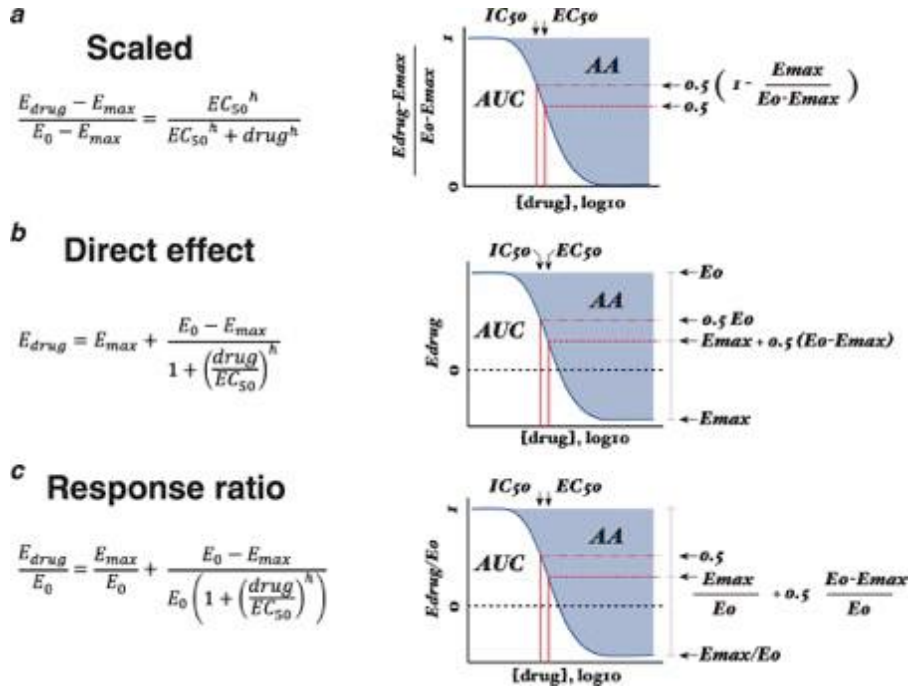

**Supplementary Figure 1 | *Different formulations of the dose–response curve***
Potency parameters $EC_{50}$ and $IC_{50}$ are shown, as are area under the curve (AUC) and activity area (AA; the inverse of AUC), parameters that attempt to capture both potency and efficacy in a single quantity. **(a)** The "scaled" form given in equation (S2); **(b)** The "direct effect" form obtained by rearranging equation (S2) to solve for $E_{drug}$; **(c)** The "response ratio" form obtained by dividing the direct-effect form by $E_0$. Note that we consider here a case where $E_{max} < 0$, which is possible when using a dynamic drug effect metric such as DIP rate. This results in $IC_{50} < EC_{50}$ (see equation S5). In cases like this, the $IC_{50}$ is sometimes referred to as the $GI_{50}$ (half-maximal growth inhibitory concentration; see Supplementary Table 1).

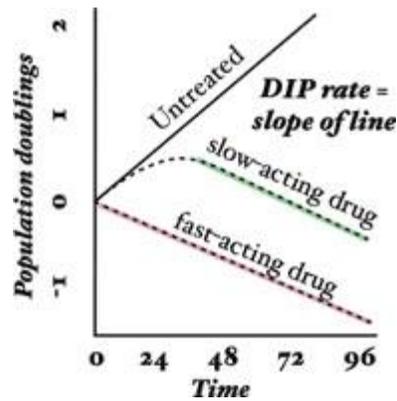

**Supplementary Figure 2 | *Fast-acting drugs, slow-acting drugs, and DIP rate***
Hypothetical growth curves (in log scale) for a cell line untreated and treated with two different drugs: a fast-acting drug where the full effect is achieved immediately, and a slow-acting drug that causes a temporal delay in the stabilization of the drug effect. Also shown is drug-induced proliferation (DIP) rate, defined as the slope of the line after the drug effect has stabilized (in this case, immediately for the fast-acting drug and ≥48h for the slow-acting drug). Note that the DIP rate is shown as equivalent for both the fast- and slow-acting drugs for illustration purposes only.

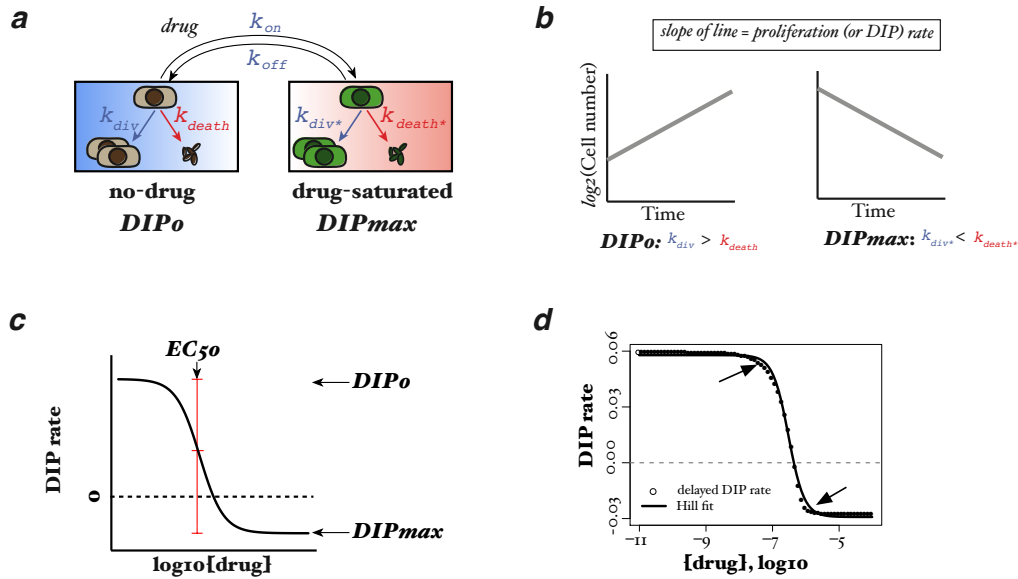

**Supplementary Figure 3 | *Two-state model of fractional proliferation predicts a sigmoidal relationship between proliferation rate and drug concentration.***
**(a)** The model assumes two states, a drug-naïve state and a drugged state, each with its own characteristic rate of proliferation ($DIP_0$ and $DIP_{max}$, respectively), which is the difference between the rates of cell division and death. The rate of transition from the drug-naïve state to the drugged state depends on the concentration of drug, while the reverse transition does not. Hence, as the concentration of drug increases, the dynamic equilibrium between states shifts increasingly in favor of the drugged state. **(b)** Since the action of an antiproliferative drug is to reduce, and perhaps reverse, the rate of proliferation of a cell population, we assume that the proliferation rate of the drug-naïve state is positive and greater than that of the drugged state (which may be positive or negative). In Figure 1 of the main text, we assume that in each case the drug is cytotoxic at saturating drug concentrations (i.e., causes regression of the cell population). Hence, the DIP rate of the drugged state ($DIP_{max}$) is assumed to be negative. **(c)** An example dose–response curve predicted by the two-state model under the partial equilibrium assumption (PEA). The curve was generated from equation (S27) with $EC_{50}$ = 1e–8 M, $DIP_0$ = 0.06*ln(2) h$^{-1}$, and $DIP_{max}$ = –0.03*ln(2) h$^{-1}$. **(d)** An example dose–response curve predicted by the two-state model in conditions where the PEA does not hold. The curve was generated by numerical integration of equations (S12) and (S13) with $k_{on}$ = 1e5 M$^{-1}$ h$^{-1}$, $k_{off}$ = 1e–3 h$^{-1}$, $k_{div}$– $k_{death}$ = 0.06*ln(2) h$^{-1}$, and $k_{div*}$ – $k_{death*}$ = –0.03*ln(2) h$^{-1}$. Note that these are consistent with the values used in part (c); see equations (S22), (S23), and (S26). Arrows highlight largest differences between calculated values (circles) and the Hill equation fit (black line).

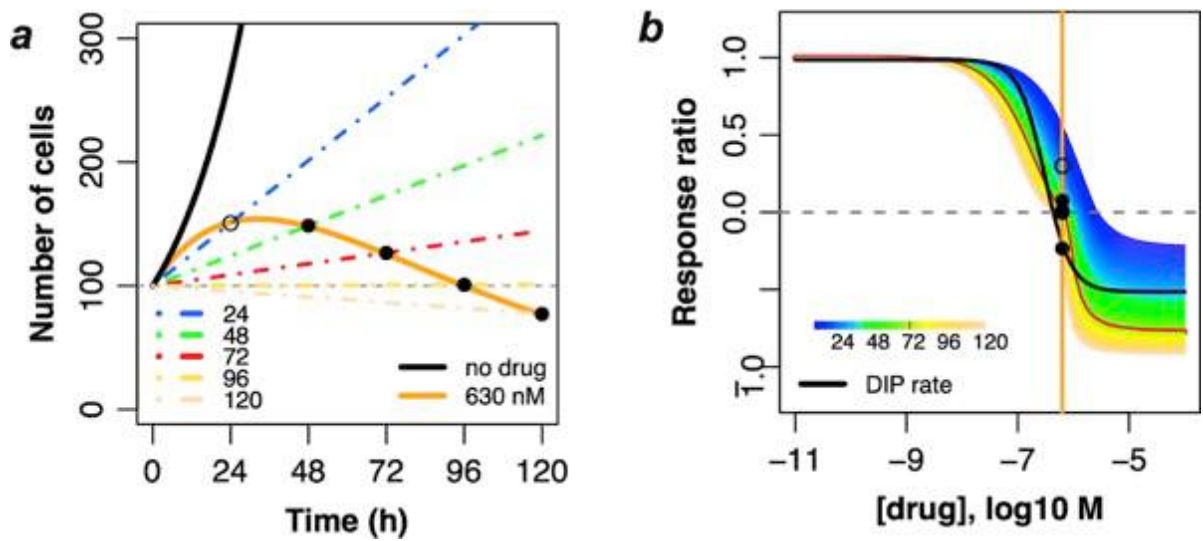

**Supplementary Figure 4 | *Theoretical illustration of bias in dose–response curves based on the NCI DTP dynamic drug effect metric***

**(a)** Growth curves in the absence and presence of 630 nM drug for the theoretical fast-growing cell line with delayed drug effect (Fig. 1c in the main text). Dash-dotted lines are a visual illustration of the NCI DTP dynamic metric and the time-dependent bias that it harbors. Depending on when cell count measurements are taken, the NCI DTP metric can indicate that the drug is partially cytostatic (<96h), fully cytostatic (96h), or cytotoxic (>96h) at this concentration. **(b)** Comparison of dose–response curves for this cell line and drug type based on the NCI DTP dynamic effect metric and DIP rate. The vertical orange line corresponds to 630 nM drug concentration; circles correspond to those in part (a).

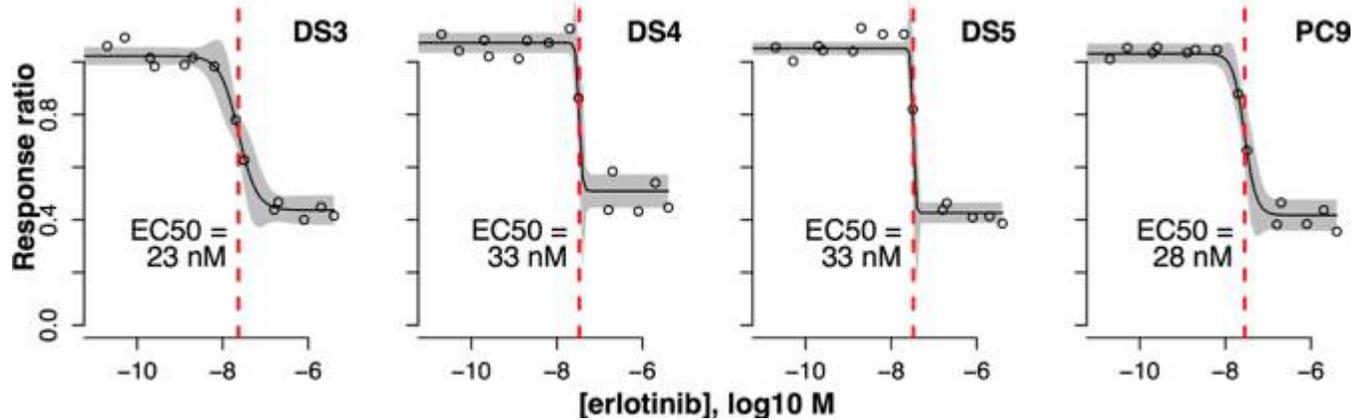

**Supplementary Figure 5 | *Static-based dose–response curves for PC9 parental cells and subclones treated with erlotinib***

Dose–response curves for three PC9-derived drug-sensitive (DS) clonal sublines (DS3, DS4, DS5) and parental PC9 cells treated with erlotinib using cell counts after 72h drug exposure as the drug effect metric. Filled circles are mean values ($n \geq 6$), lines are optimal fits to a four-parameter log-logistic model (equation S2), and grey shading indicates 95% confidence intervals on the fitting function. None of the dose–response curves for the clonal sublines is statistically different from the parental curve based on *t*-statistics and the null hypothesis that the ratio of clonal to parental values is 1 ($p > 0.05$ for each of the four fitting parameters).

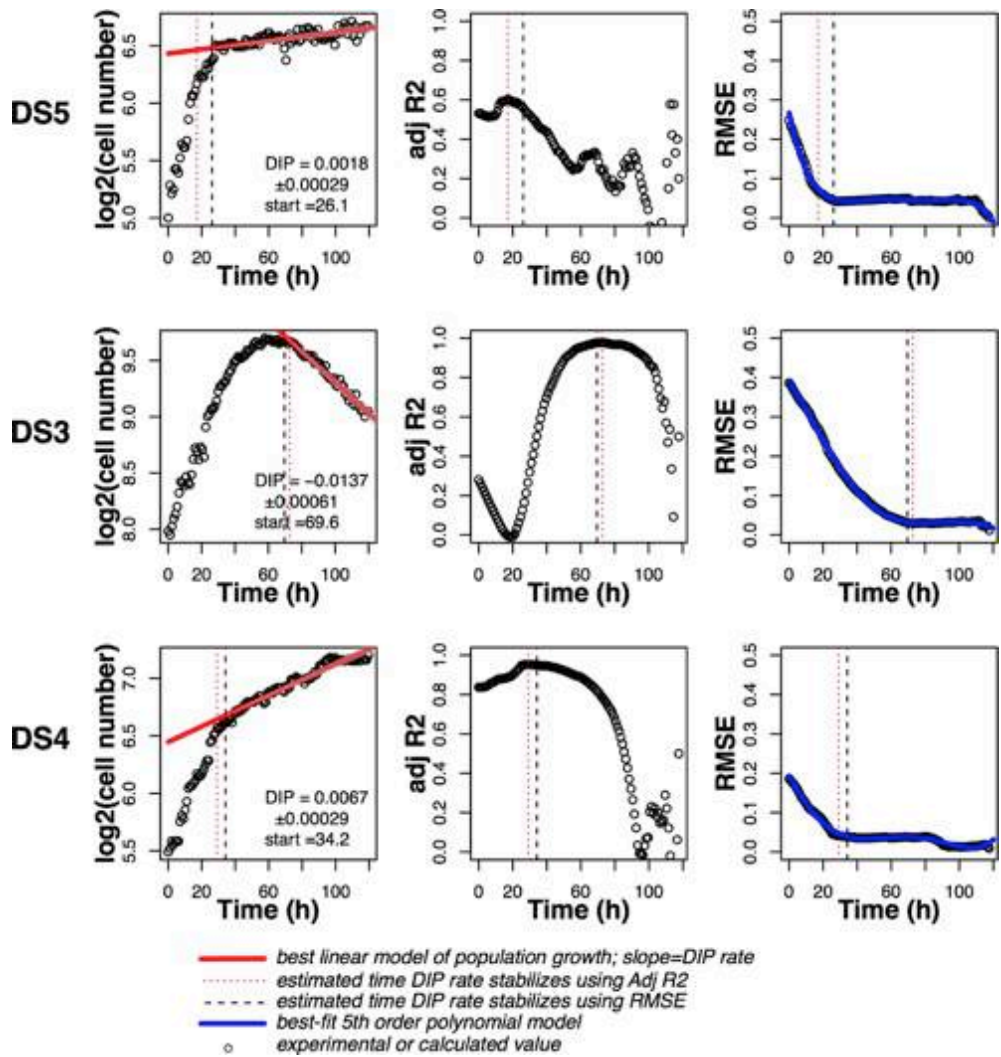

**Supplementary Figure 6 | *Automated estimation of DIP rate from cell count data***
Cell count data for the three PC9 sublines from Figure 2c of the main text is used to evaluate
the ability to automatically estimate drug effect stabilization times and DIP rates using adjusted
$R^2$ and root-mean-squared error (RMSE) as measures of linearity. Best linear model fits to the
population growth curves (solid red lines) are based on the RMSE-estimated stabilization times.
A fifth-order polynomial (solid blue line) was fit to the RMSE curve and used to estimate the
point at which exclusion of additional data points does not substantially improve the linear model
fit. Source code implementing this approach (`dipDRC.r`) is freely available at
github.com/QuLab-VU/DIP_rate_NatMeth2016. See Supplementary Note for more details.

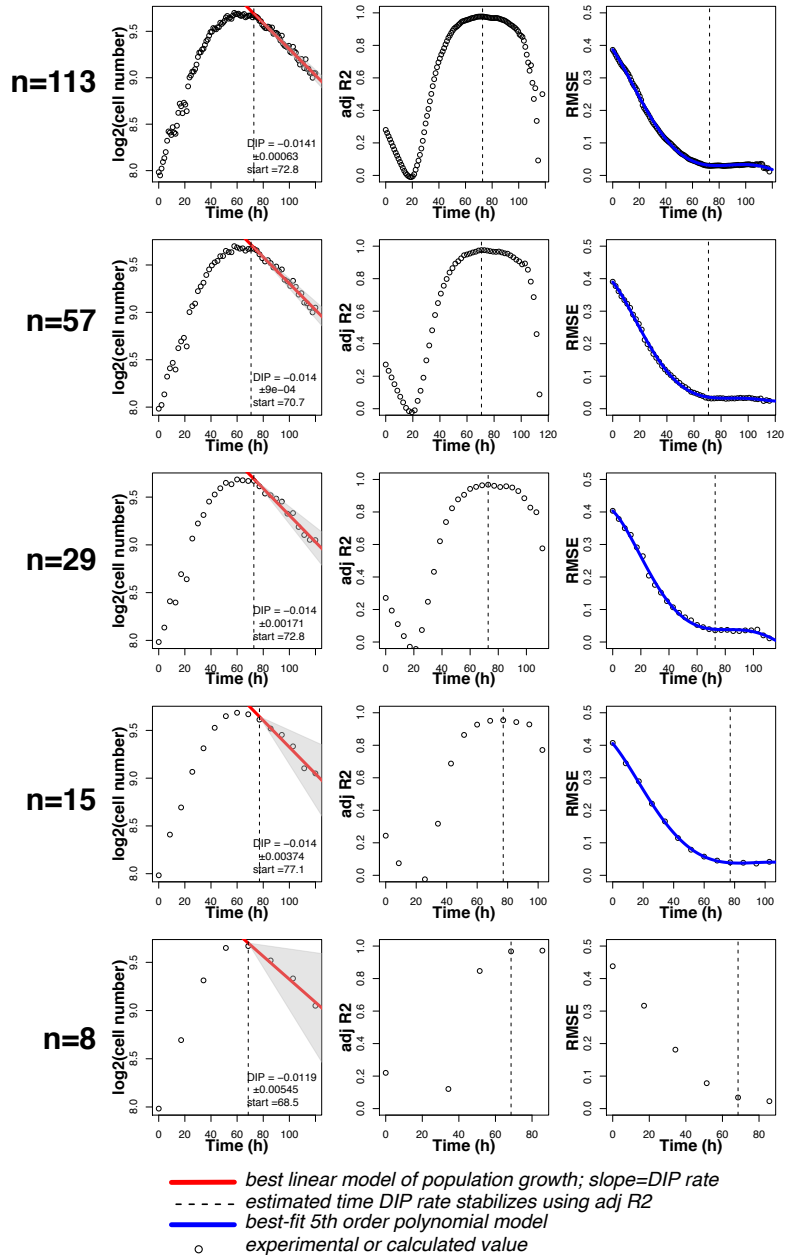

best linear model of population growth; slope=DIP rate
estimated time DIP rate stabilizes using adj R2
best-fit 5th order polynomial model
experimental or calculated value

**Supplementary Figure 7 | *Effects of sampling frequency on automated DIP rate estimation***

Cell count data for one PC9 subline (DS3) from Figure 2c of the main text is used to evaluate the robustness of the automated DIP rate estimation method to changes in sampling frequency. The full data set (top row) was successively subsampled a total of four times. Best linear model fits (solid red lines) are based on the adjusted $R^2$-estimated stabilization times. DIP rates for data sets with ≥15 data points varied by less than 1%; DIP rate from the smallest data set (n=8) was within 15% of that for the full data set. Gray shading indicates 95% confidence interval on the DIP rate. See Supplementary Note and Supplementary Figure 6 for more details.

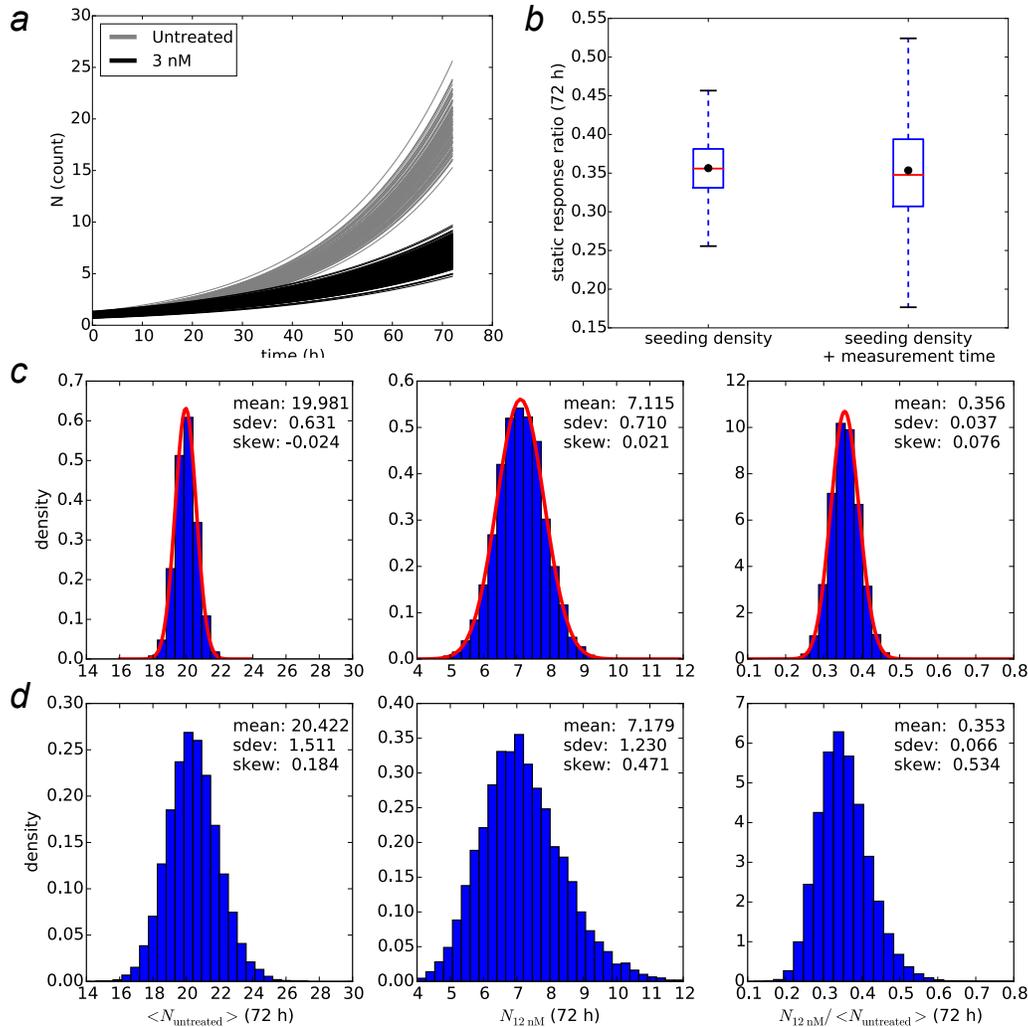

**Supplementary Figure 8 | *Theoretical effects of variations around a mean cell seeding density on the static effect metric and response ratio***

The two-state model of fractional proliferation (Supplementary Note and Supplementary Fig. 3) with the "fast proliferating / fast acting" parameter set (Supplementary Table 2) is seeded with initial cell counts drawn from a normal distribution with mean 1 and standard deviation 0.1 and simulated under untreated and drug-treated conditions. **(a)** 100 simulated time courses each of untreated and drug-treated cell populations. **(b)** Boxplots of static response ratios at 72h for variations in seeding density alone and variations in seeding density and measurement time (72 ± 5h). Based on $10^8$ samples. Red line is the median; black dot is the mean; boxes extend from the first to third quartile; whiskers extend 1.5 times the interquartile range; outliers are not shown. **(c)** For variations in the seeding density alone, histograms for the mean of 10 untreated final (72h) cell counts (n=$10^4$; *left column*), the drug-treated final cell count (n=$10^4$; *middle column*), and the static response ratio (n=$10^8$; *right column*). Sample means, standard deviations, and skews are shown in each case. Analytical distributions are shown in red (equations S30, S32, and S33). **(d)** Same as *c* (less the analytical distributions) but for variations in both seeding density and measurement time (72 ± 5h).

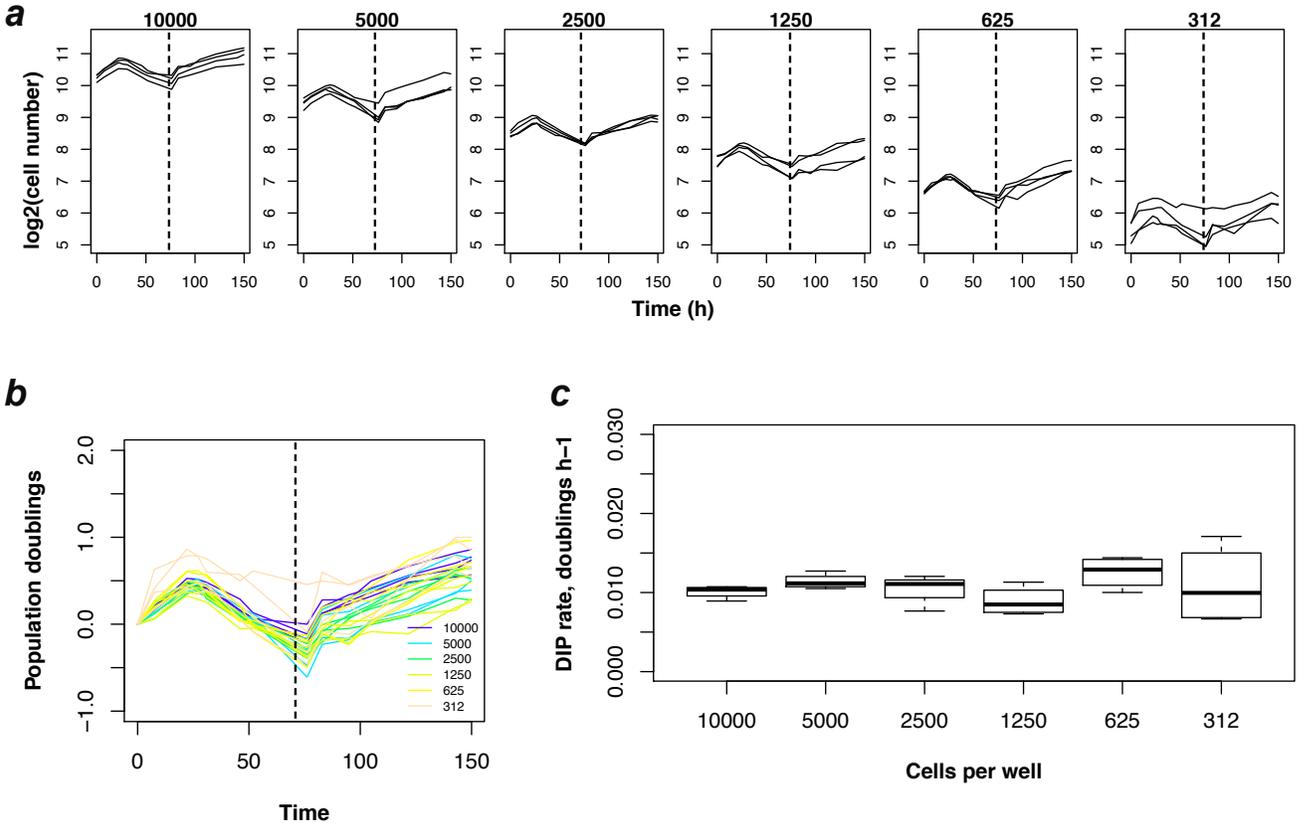

**Supplementary Figure 9 | *Effects of variations in mean cell seeding density on DIP rate estimation***

Cell count data for BRAF-mutant (SKMEL5) melanoma cells treated with 8 $\mu$M PLX4720 is used to evaluate the effects of mean seeding density on DIP rate estimates. Cell counts were obtained using fluorescence microscopy imaging (Online Methods). **(a)** Population growth curves ($\log_2$ scaled) for each seeding density considered (n=4; seeding density is listed above each plot). Vertical dashed line corresponds to ~72h, the hand-chosen stabilization time; data to the right of this point were used to estimate DIP rate. **(b)** Population growth curves from part (a) normalized to the number of cells for each well at the first time point. Vertical dashed line corresponds to ~72h. **(c)** Boxplots of estimated DIP rates (n=4) at each seeding density. Mean values were statistically indistinguishable across all seeding densities (p=0.47); variances were not (Levene's test[25], p=0.0082), as expected (see Supplementary Note).

**Supplementary Table 1 | *Glossary of terms***

| | |
|---|---|
| *Activity area (AA)* | The area above a dose–response curve between the upper ($E_0$) and lower ($E_{max}$) plateaus[3]; a drug-activity parameter that attempts to quantify both potency and efficacy in a single value; the inverse of the AUC. |
| *Area under the curve (AUC)* | The area below a dose–response curve between the upper ($E_0$) plateau and a defined lower bound (not necessarily the lower plateau $E_{max}$); a drug-activity parameter that attempts to quantify both potency and efficacy in a single value; the inverse of the AA. |
| *Drug activity* | General term for the action of a drug on cells; in assays of cellular proliferation, refers to the effect that a drug has on cell division and death (e.g., slowing the rate of progress through the cell cycle; triggering the apoptotic machinery). |
| *Drug-activity parameter* | Any quantified value that can be extracted from a dose–response curve (e.g., $E_{max}$, $EC_{50}$, $GI_{50}$, $h$, $IC_{50}$, AA, AUC). |
| *Drug effect* | The induced response of a population of cells to a drug. |
| *Drug effect metric* | A measured value that quantifies drug effect (e.g., cell number after a specified period of drug exposure, DIP rate). |
| *Drug-induced proliferation (DIP) rate* | The steady-state rate of expansion or regression of a cell population; defined in terms of population doublings ($\log_2$-scaled cell numbers) per unit time; estimated as the slope of the line on a plot of population doublings vs. time after the drug effect has fully stabilized; a proposed metric of antiproliferative drug effect *in vitro*. |
| *Dynamic metric* | A metric whose value is based on measurements at two or more time points (e.g., DIP rate, NCI DTP metric). |
| $E_0$ | The quantified effect in the absence of drug. |
| $E_{drug}$ | The quantified effect at a given concentration of drug. |
| $E_{max}$ | The quantified effect at saturating drug concentrations; a measure of efficacy[1]. |
| $EC_{50}$ | The concentration of drug at which the effect is halfway between the minimum ($E_0$) and maximum ($E_{max}$) effects; a measure of potency[26]. |
| *Efficacy* | The degree to which a drug can produce a beneficial effect. |
| *Fractional proliferation* | The concept that at a given drug concentration, cells experience multiple fates (e.g., division, death, entry into quiescence) and that the cell population dynamics are a combined effect of all of these events[8]. |
| $GI_{50}$ | The concentration of drug at which the effect is 50% of untreated control ($E_0$/2), independent of $E_{max}$; technically equivalent to $IC_{50}$, the term $GI_{50}$ is often used for dose-response curves where $E_{max}$ < 0 (refs [1] and [11]); a measure of potency. |
| $h$ | Hill coefficient; proportional to the slope of the dose–response curve at $x = EC_{50}$; shown to correlate with the degree of heterogeneity intrinsic to a cell line[1]. |
| $IC_{50}$ | The concentration of drug at which the effect is 50% relative to untreated control ($E_0$/2), independent of $E_{max}$; a measure of potency[1,4,27,28]. |
| *Potency* | The amount of drug required to produce a specified effect; a highly potent drug is active at low concentrations. |
| *Response ratio* | Ratio of the quantified effect at a given drug concentration ($E_{drug}$) to the effect in the absence of drug ($E_0$); often used as the dependent variable in dose–response curves. |
| *Static metric* | A metric whose value is based on a measurement at a single time point (e.g., cell number after a specified period of drug exposure). |
| *Time-dependent bias* | Degree to which the value of a metric varies with the chosen measurement time point(s). |
| *Unbiased metric* | A metric whose value is independent of the chosen measurement time point(s) (e.g., DIP rate). |

**Supplementary Table 2 |** *Rate parameter values used in this work*

| Parameter | Theoretical cell type / drug type | | |
|---|---|---|---|
| | *Fast-proliferating / Fast-acting* | *Slow-proliferating / Fast-acting* | *Fast-proliferating / Slow-acting* |
| $k_{div} - k_{death}$ (h$^{-1}$) | 0.06*ln(2) | 0.01*ln(2) | 0.06*ln(2) |
| $k_{div^*} - k_{death^*}$ (h$^{-1}$) | −0.03*ln(2) | −0.005*ln(2) | −0.03*ln(2) |
| $k_{on}$ (M$^{-1}$ h$^{-1}$) | 1e8$^\dagger$ | 1e8$^\dagger$ | 1e5 |
| $k_{off}$ (h$^{-1}$) | 1$^\dagger$ | 1$^\dagger$ | 1e−3 |

$^\dagger$These values were not set independently but as the ratio $k_{off}/k_{on}$ (see equation S20).

**SUPPLEMENTARY REFERENCES**